\documentclass[
    ,final            
  ]
  {aipproc}

\layoutstyle{6x9}

\def\beq{\begin{equation}}   \def\eeq{\end{equation}}
 
\def\bea{\begin{eqnarray}}   \def\eea{\end{eqnarray}}

\newcommand{\MeV}{\,\mbox{MeV}}

\newcommand{\gsim}{\lower.7ex\hbox{$ \;\stackrel{\textstyle>}{\sim}\;$}}
\newcommand{\lsim}{\lower.7ex\hbox{$ \;\stackrel{\textstyle<}{\sim}\;$}}

\def\epem{ e^+e^- }
\def\c2{CLEO~II.V}

\def\cqb{ c\bar{q} }
\def\csb{ c\bar{s} }
\def\bsb{ b\bar{s} }
\def\bqb{ b\bar{q} }
\def\d0d0{ D^0\bar{D}^0 }
\def\p0p0{ P^0\bar{P}^0 }

\def\qp2{ \Bigl| \frac{q}{p} \Bigr|^2 }
\def\pq2{ \Bigl| \frac{p}{q} \Bigr|^2 }

\def\ps2s{  \psi(2S) }
\def\q2{ $q^2$ }
\def\cm2s1{ $\,{\rm cm}^{-2} {\rm s}^{-1}$} 
\def\DsJ{ D_{sJ} }

\begin{document}

\title{Spectroscopy of D Mesons}


\author{Stefano Bianco}{
  address={Laboratori Nazionali di Frascati dell'INFN \\ v.E.~Fermi
  40, 00044 Frascati (Rome) Italy}
}

\begin{abstract}
 The scenario of heavy quark meson spectroscopy underwent recently a
 major revolution, after the observation of BABAR and CLEO, confirmed
 by BELLE, of $\DsJ$ L=1 excited states, and by further evidences by
 SELEX. These experimental results have cast doubts on the incarnations of the
  ideas of Heavy Quark Effective Theory in heavy quark spectroscopy. I shall review the
  status of  experimental data,  discuss implications and sketch an outlook.
\end{abstract}

\maketitle

%
\section{Introduction}
This paper reports on recent experimental results on D meson spectroscopy,
discussing the recent events that brought to cast doubts to our
current understanding of the overall picture. I shall discuss excited
non-strange D   mesons, namely the observation of $j_q=1/2$ broad
states,  the revolutionary observations of excited strange $\DsJ$ mesons
which are forcing us to switch the paradigm of HQ spectroscopy,
discuss the status of debated $\DsJ$(2632) meson observed by SELEX at
Fermilab, finally sketch an outlook and draw conclusions. For a detailed review
on charm physics including spectroscopy the reader is referred to
Ref.\cite{Bianco:2003vb}, for other charm spectroscopy issues such as charmonium
states etc. see other up-to-date reviews such as \cite{Rosner:2005gf,Seth:2005,Barnes:2005zy}.
\par
Let  me pay a tribute to cosmic ray physicists and show  the ---
possibly --- very first D meson
observed by human eye $(D^+ \rightarrow K^+ \pi^0)$, in nuclear emulsions
exposed to cosmic rays in 
1971 \cite{NIU}. After 35 years, here is where we are. 
 \par
 \begin{figure}
   \includegraphics[width=12.0cm]{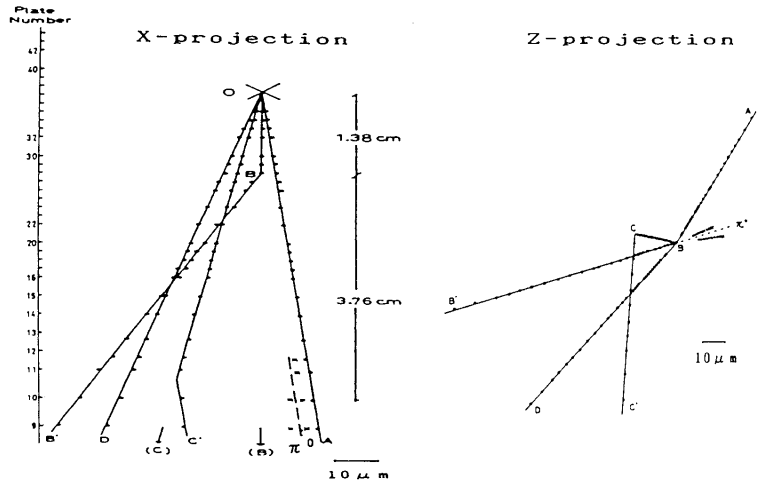} 
 \caption{First charm candidate event in nuclear  emulsions \cite{NIU}. 
 Figure from Ref.~\cite{NIUNARRATED}. 
 \label{FIG:NIU}}   
\end{figure}
   \section{Heavy-Light Quark Spectroscopy, The Global Picture}
A global interpretation scheme for heavy quark meson spectroscopy is provided
by the idea of Heavy Quark Symmetry (HQS). In the infinite heavy-quark mass
limit, the heavy-light meson can be described as formed by a the still heavy
quark, with all the orbital degrees of freedom being due to the light quark.
This means that good conserved quantum numbers are the spin of the 
heavy quark, and the angular momentum $j_q$.  
\par
\begin{figure}
  \includegraphics[width=10.0cm,height=7.0cm]{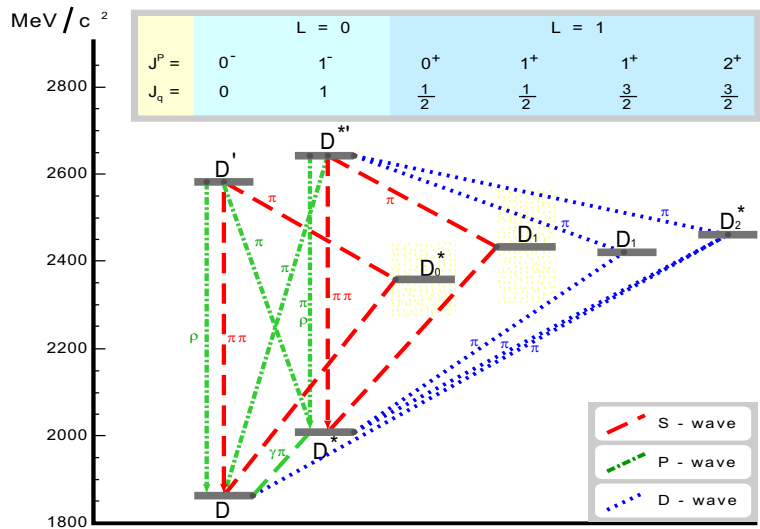}
 \caption{\it
  Masses and transitions predicted for the excited non-strange D meson states.
    \label{FIG:DSPEC} }
\end{figure}

Experimentally, for each of the $c\bar u$, $c\bar d$ and $c \bar s$ systems
four  P-wave and two $n=2$ radial excitations have been studied.
There are four $L=1$ states, namely two with $j_q=1/2$ and total spin
$J=0,1$ and two with $j_q=3/2$ and $J=1,2$. These four states
are named respectively $D_0^*$, $D_1(j_q=1/2)$, $D_1(j_q=3/2)$ and
$D_2^*$ (Fig.\ref{FIG:DSPEC}). 
Parity and angular momentum conservation force the $(j_q=1/2)$ states to
decay to the ground states via S-wave transitions (broad width), while 
$(j_q=3/2)$ states decay via D-wave (narrow width). To be more specific, for
the  $1/2$ one predicts widths of $\sim 100$~MeV and for the $3/2$ of about
$\sim 10$ MeV with the exception of the  $D_{s1}(j_q=3/2)(2536)$ which is
kinematically forced to a $\sim 1$~MeV  width. 
\par
Therefore, the HQS picture has two consequences, which turn to be direct
predictions: 
\begin{enumerate}
 \item each L level is split in two J-degenerate doublets, in each doublet one
 broad and one narrow state;
 \item flavour symmetry does exist. In principle in the heavy-quark infinite
 mass limit one is allowed to use the same chart tfor $ \cqb , \csb , \bqb ,
 \bsb $ mesons, just changing quark labels and absolute  mass energy scale.
\end{enumerate}
However,  the HQS paradigm was recently put in discussion by BABAR's and CLEO's
discovery of $\DsJ$ states.
\section{L=1 non-strange excited D mesons}
All six $L=1, j=3/2$  non-strange {\em narrow} states are well established, 
 with precisions on masses at the 1 MeV level and on widths at the few MeV
 level. This is due to the fact that excited D states are abundantly produced
 both at FT experiments, in $\epem$ continuum production, in B decays and at
 the  $Z^0$ \cite{Roudeau:1997zn}.
 \par
 Table~\ref{TAB:DSTCQ} shows the experimental data available for $\cqb$ L=1 mesons,
 masses and widths, as showing on Ref.\cite{Eidelman04} updated to 2005, as well as recent 
 measurements not appearing in PDG world averages. In bold I listed measurements
 that are somehow new or debated.  
\par
 Let me first of all mention a long-standing dilemma, the $D^{*\prime}$.  Called
 $D^*(2640)^\pm$ by PDG, the first L=1 radial excitation was seen by
 DELPHI \cite{Abreu:1998vk}  in the $D^{*+}\pi^-\pi^+$ final state;
 it has  not been confirmed by any experiment  
   (OPAL\cite{Abbiendi:2001qp},   CLEO\cite{Rodriguez:1998ng},
  ZEUS\cite{Sefkow:2000eu}). Final disproof or confirmation is needed, and it should be
  considered as a relatively easy task considered the level of statistics currently
  available to contemporary experiments.
 \par 
The status of the {\em broad} L=1  states is not clear at all, as well. The
assignments of the quantum numbers are largely based on theory 
expectations for their masses and widths. 
In 1998 CLEO \cite{Anderson99}  showed evidence for the $D_1(j_q=1/2)$ broad state. 
Two  results, by BELLE \cite{AbeDst} and photoproduction experiment FOCUS
\cite{LinkDst}, have appeared in 2003 and are now included in the average of PDG
2005. BELLE have studied the $D^{*+}\pi^-$ and $D^+\pi^-$ final states, while FOCUS
have studied both isospin channels  $D^+\pi^-$ and $D^0\pi^+$. They both claim
observation  for broad states. Due to the presence of feeddown satellite peaks due
to missing neutral kinematics, FOCUS do not claim conclusively that the broad state observed is
the $D^{*}_0$ predicted by HQS. The mass values found are in disagreement at the
$\sim 2\sigma$ level, and consistent with many predictions out of the huge
number of papers on the subject. The BELLE mass value is notably close to
what predicted a long time ago \cite{Matsuki:1997da}.
 More experimental results are needed. 
\par
 New players in the D meson spectroscopy game could be the experiments at hadron
 colliders, which have greatly improved charm physics capabilities with impact
 paramenter trigger which uses silicon vertex detectors.
 As instance, CDF at the Fermilab Tevatron showed results in 2003
 \cite{Shapiro03} with high statistic peaks of L=1 mesons sitting on huge 
 combinatoric backgrounds, due to high multiplicity of primary interaction vertex.
 Clearly, hadroproduction is not the best place to look for L=1 mesons. However,
 more recent unpublished results \cite{CDF7191} show great improvements, with $D\pi$
 distributions clearly evidencing clean L=1 mesons peaks. We expect interesting news from
 CDF and D0, possibly at this same conference.
\par
%
%
\par
\begin{table}
 \caption{ Summer 2005 status of (L=1, n=1) and (L=0, n=2)   $c\bar q$ mesons (MeV).
  Statistical and  systematical errors added in quadrature, unless noted. Preliminary data
  from CDF   are from \cite{Shapiro03}\cite{CDF7191}.
  \label{TAB:DSTCQ} 
 }
 \begin{tabular}{|l|c|c|c|c|c|c|} \hline
 $j_q$     & $1/2$    & $1/2$  & $3/2$ & $3/2$   & $1/2$ & $1/2$  \\
 $J^P$     & $0^+$    & $1^+$  & $1^+$ & $2^+$   & $0^-$ & $1^-$  \\
 $L,n$     & $1,1$    & $1,1$  & $1,1$ & $1,1$   & $0,2$ & $0,2$  \\
\hline
 & $D_0^*(2400)$ & $D_1(2430)$ & $D_1(2420)$   & $D_2^*(2460)$
 &$D^\prime$&$D^{*\prime}\,[D^*(2640)]$ 
 \\ 
  Decay Mode                &
    $D\pi$                  &
    $D^*\pi$                &
    $D^*\pi$                &  
    $D\pi,D^*\pi$           &
                            &
  $D^*\pi\pi$  \\      
\hline
 \multicolumn{7}{|c|}{Mass (MeV)}    \\
   PDG $0$            &          
   {\bf 2352 $\pm$ 50}          &
    2427 $\pm$ 36               &  
    2422 $\pm$  2               &
   {\bf 2461.1 $\pm$ 1.6}       &  
                                &
                                 \\ 
 PDG  $\pm$                         &          
    2403 $\pm$ 38               &
                                &  
    2427 $\pm$  5               &
 {\bf  2464.9 $\pm$ 3.0 }               &  
                                &
    2637 $\pm$  7    \\
  CDF  prel. $\pm$                          &          
                   &
                                &  
                  &
 {\bf  2463.6 $\pm$ 2.7 $\pm$ x }               &  
                                &
       \\      
         CDF  prel. $0$                          &          
                   &
                                &  
        {\bf  2421.7 $\pm$ 0.9 }          &
 {\bf  2463.3 $\pm$ 1.0 }               &  
                                &
       \\		          
	  \hline 
 \multicolumn{7}{|c|}{Width (MeV)}    \\	  
 PDG  $0$      &          
    {\bf  261 $\pm$ 50 }  & 
	 384 $\pm$ 117        &
	  19 $\pm$ 4        &  
     {\bf 32 $\pm$ 4 }    &
	                    &
	                  \\  
 PDG $\pm$                        &          
        283 $\pm$ 42                     & 
	                     &
	  28 $\pm$ 8        &  
          29 $\pm$ 5     &
	                    &
	     $<$ 15           \\ 
         CDF  prel. $0$                          &          
                   &
                          &  
        {\bf  20 $\pm$ 2 }          &
 {\bf  49 $\pm$ 3 }               &  
                                &
       \\		      		       
      \hline
 \multicolumn{7}{|c|}{Isospin Mass Splitting(MeV)}    \\	        
 PDG                         &          
                        & 
	                     &
 {\bf $4^{+2}_{-3}\pm 3$}        &  
        {\bf 2.4  $\pm$ 1.7}     &
	                    &
	      \\
 \hline
 \end{tabular}
\end{table}
\section{L=1 strange excited D mesons, or: Need to Change Paradigm of D
Spectroscopy ?} 
Before Spring 2003 we thought we could use the same $\cqb$ chart in Fig.\ref{FIG:DSPEC}
 for $\csb$, thanks 
to flavour symmetry of HQS. The narrow $D_{s1}$
and $D_{s2}^*$ states have been very well established since a long time, and we would expect
the two missing broad $\csb$ states to lie somewhere above the $DK$ and $D^*K$ threshold,
respectively. 
\par
Instead, surprisingly enough, BABAR finds\cite{BABARDS**} a prominent peak at
2317\MeV in $D_s\pi^0$ 
with width compatible to experimental resolution.
They also find another narrow peak in $D_s^* \pi^0$, but are not sure whether 
it is a reflection or not, therefore do not claim observation for a second state.
The analysis is complicated by the presence of two reflections from undetected
neutrals.  Following BABAR announcement, 
CLEO looked back to circa-1995 data, at the time when they published 
\cite{Gronberg:1995qp}
the first evidence for isospin-violation decay $D_s^*\rightarrow  D_s \pi^0$. At that time
they had much less statistics, now they integrate all events and they also
confirm\cite{CLEODS**} 
the BABAR state. By  availing of a more trained analysis they find and interpret
correctly the $D_s^*\pi^0$  state at 2463\MeV        as another new state.
BELLE joins the club by finding evidence\cite{Abe:2003jk,Abe:2003vu} of the $\DsJ
(2463) \rightarrow D_s \gamma$ 
and determines the $J^{PC}$. A detailed historical account is reported in
\cite{Bianco:2003vb}.
\par
 \begin{figure}[t]
   \includegraphics[width=3.0cm]{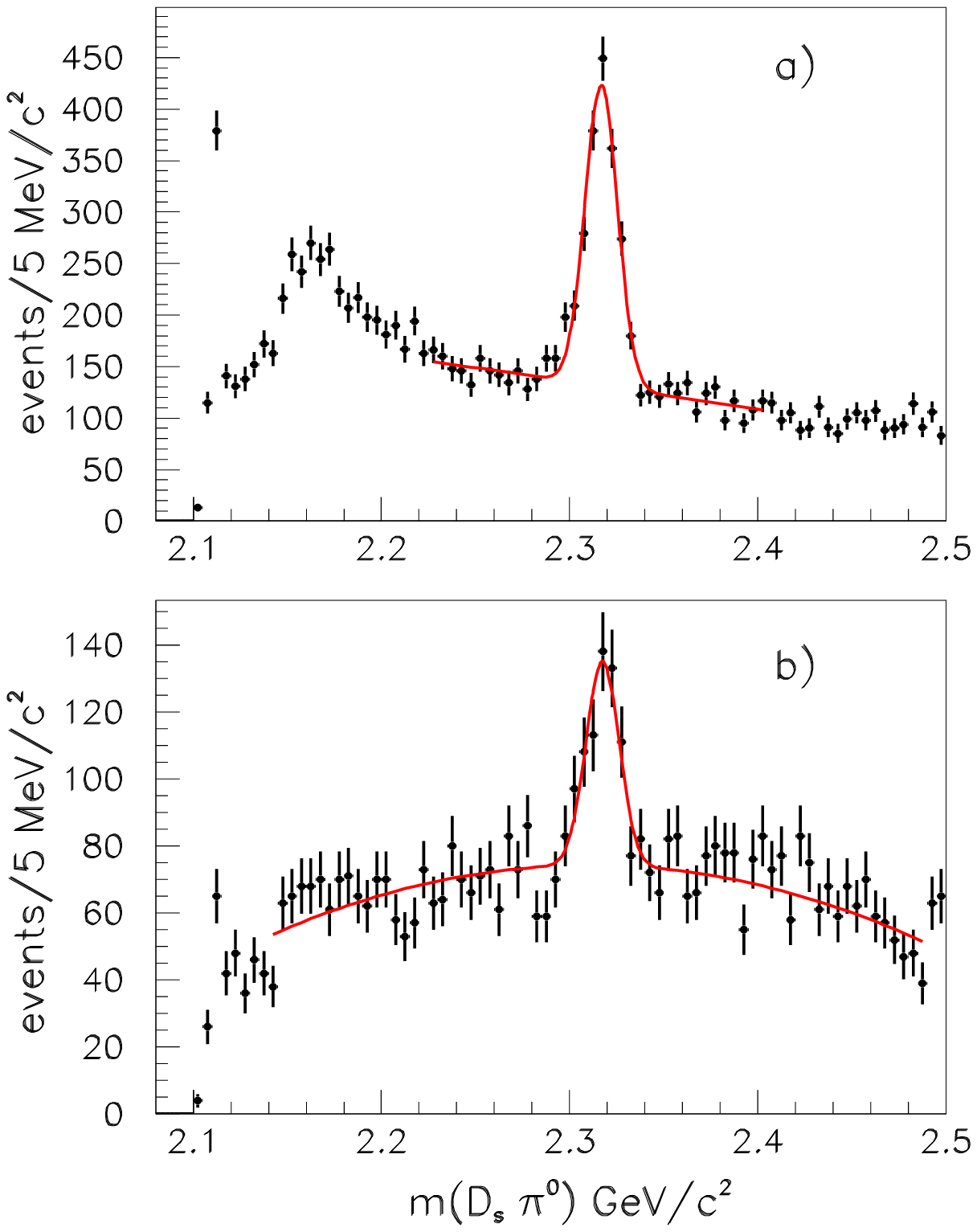}
   \includegraphics[width=3.0cm]{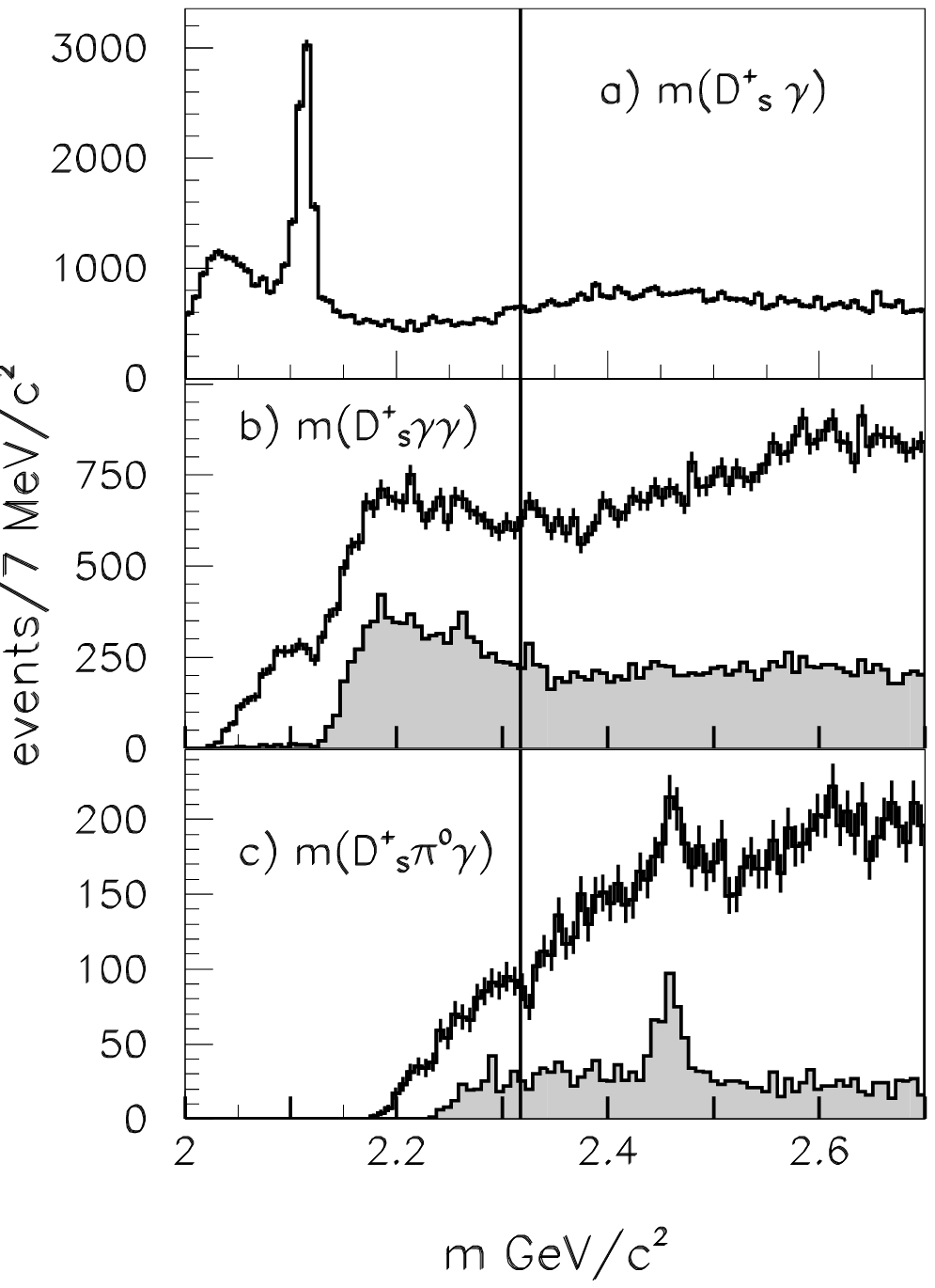}
   \includegraphics[width=3.0cm]{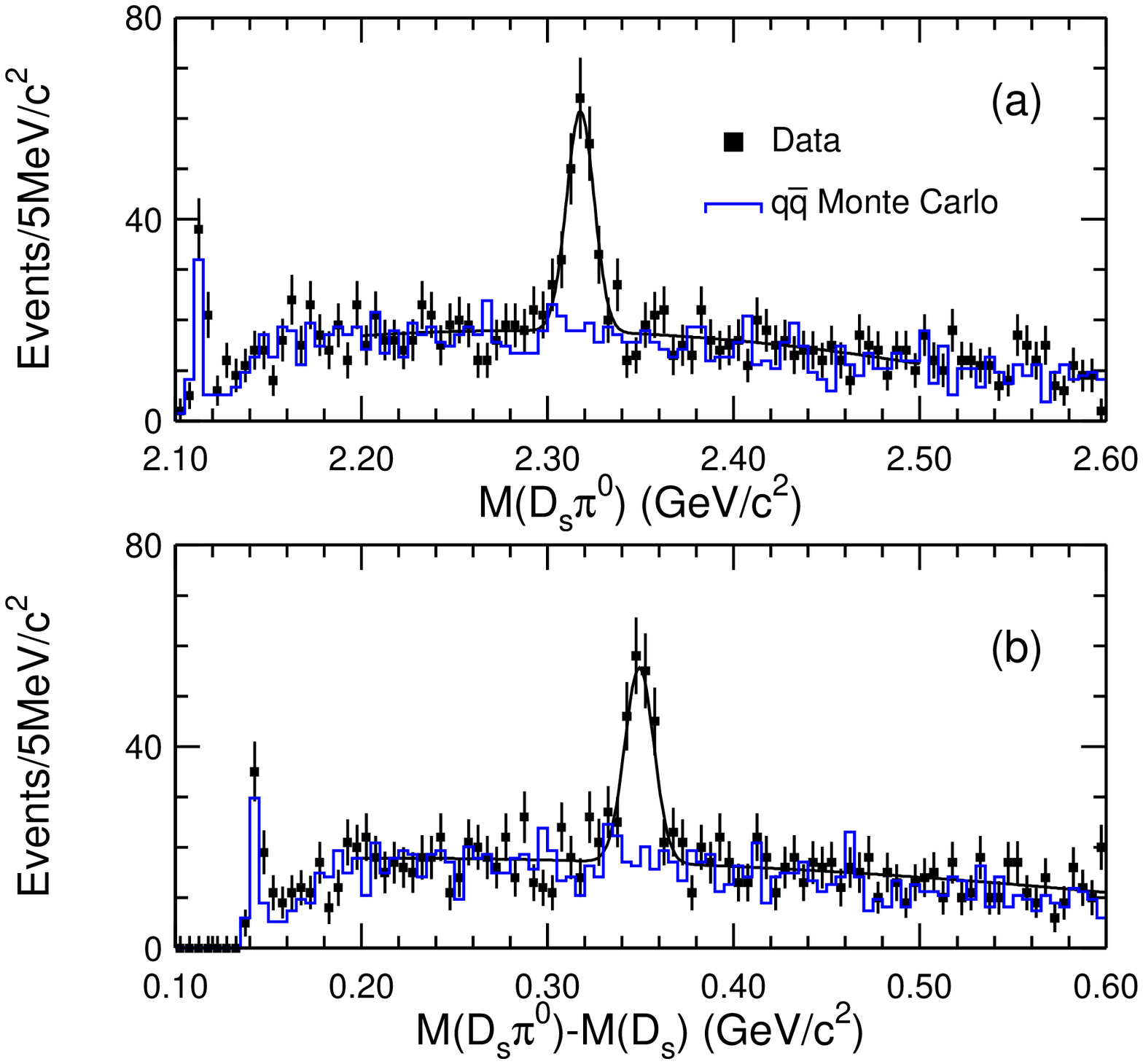}
   \includegraphics[width=3.0cm]{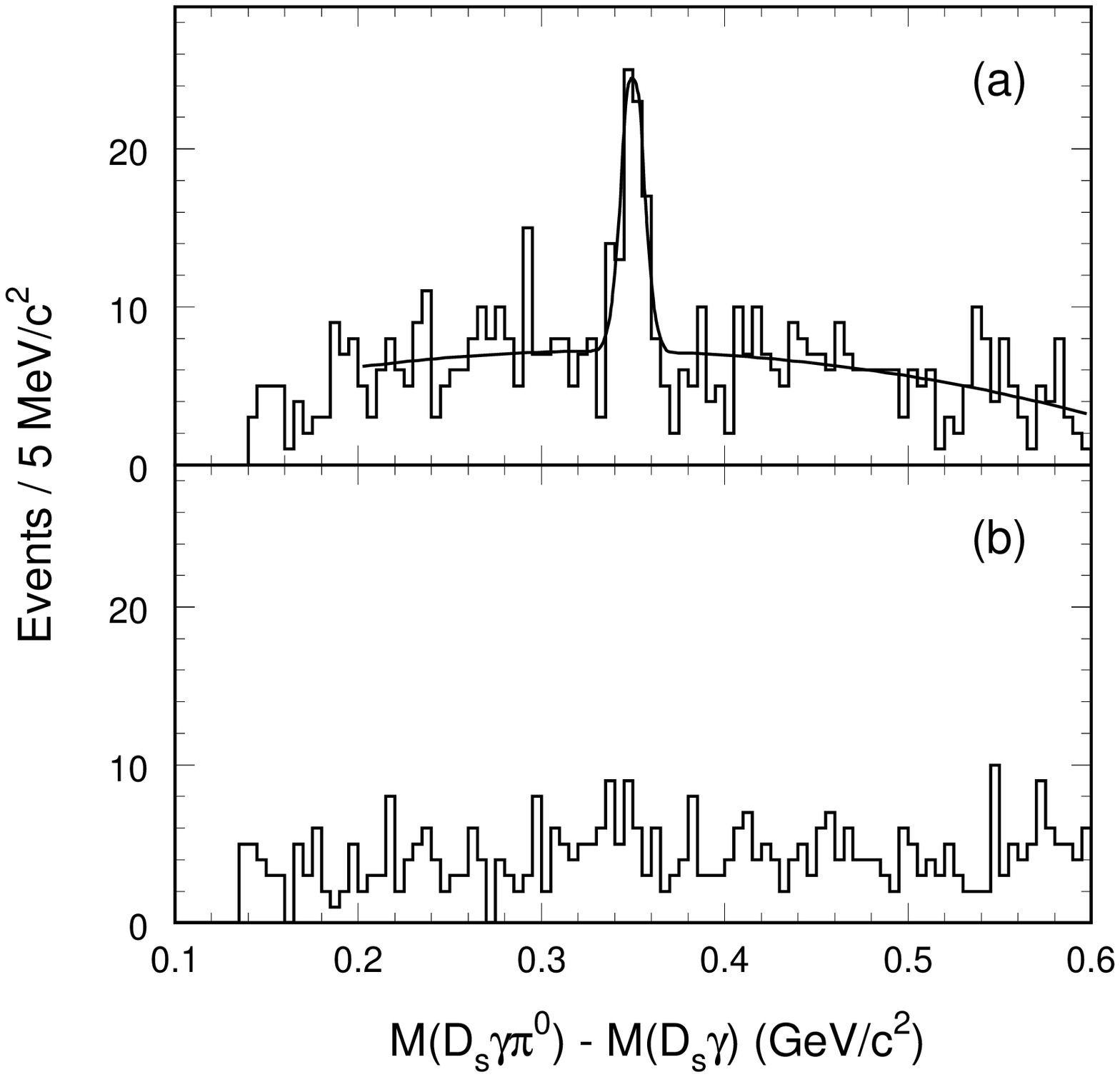}
 \caption{\it New $D^*_{s0^+}(2317)$ and $D^*_{s1^+}(2463)$ states observed by
 BABAR (a,b)  \cite{BABARDS**} and CLEO (c,d) \cite{CLEODS**}.
    \label{FIG:D0DIAG} }
\end{figure}
It seems natural to interpret $D^*_{sJ}(2317)$ and $D^*_{sJ}(2463)$ as $0^+$ and
$1^+$ states, respectively. The decay distributions are consistent with such 
assignments, yet do not
establish them. They together with the mass values would explain the narrow
widths:
for $D^*_{s1^+}(2463) \to D K$ is forbidden by parity, 
$D^*_{s0^+}(2317) \to D K$ and $D^*_{s1^+}(2463) \to D K^*$ by kinematics 
and $D^*_{s0^+}(2317) \to D_s^+ \pi^0$ and
$D^*_{s1^+}(2463) \to D_s^{*+} \pi^0$ are isospin violating transition and thus
suppressed. Also $D^*_{s0^+}(2317) \to D_s^+ \gamma$ is forbidden.
\par
There are three puzzling aspects to these states:
\begin{itemize}
\item
Why have no other decay modes been seen ? In particular CLEO places a low upper
bound
\beq
BR(D^*_{s0^+}(2317) \to D_s^{*+} \gamma ) < 0.078  \; \; 90\% \; C.L.
\eeq
Why is it not more prominent, when $D^*_{s0^+}(2317) \to D_s \pi^0$ is isospin
violating ?
\item
Why are their masses so much below predictions ? 
One should note that a deficit
of $\sim 160$ and $\sim 100$ MeV is quite significant on the scale of 
$M(D^*_{sJ}) - M(D)$. Answers to this question have been proposed a
long time ago\cite{Matsuki:1997da}.
 Why is the mass splitting to the previously found narrow
states  $D_{s1}(2536)$ and $D_{sJ}(2573)$ so much larger than anticipated ?
\item
A related mystery is the following: where are the corresponding 
{\em non}-strange
charm resonances ? They should be lighter, not heavier than $D^*_{s0^+}(2317)$
and $D^*_{s1^+}(2463)$.  
\end{itemize}
\par
PDG 2005 entries (reported in  Table \ref{TAB:DSTCS}) are dominated by the BABAR measurements.
Unpublished results not on 
PDG  are the observation of the 2317\MeV state by FOCUS, worth to be mentioned
because it is the only observation of a $D_{sJ}$ state outside $\epem$ colliders, and
some results on $1^+$ states.
\par
\begin{table}
 \caption{ Summer 2005 status of (L=1, n=1) and (L=0, n=2)  
  $c\bar s$ mesons (MeV).
 Statistical and
  systematical errors added in quadrature, unless noted.
 Preliminary results are from
 FOCUS\cite{Vaandering04}\cite{Kutschke:2000sm} 
 and BELLE\cite{BELLE0507030}.
\label{TAB:DSTCS}
 }
 \footnotesize
 \begin{tabular}{|l|c|c|c|c|c|c|} \hline
 $j_q$     & $1/2$    & $1/2$  & $3/2$ & $3/2$   & $1/2$ & $1/2$  \\
 $J^P$     & $0^+$    & $1^+$  & $1^+$ & $2^+$   & $0^-$ & $1^-$  \\
 $L,n$     & $1,1$    & $1,1$  & $1,1$ & $1,1$   & $0,2$ & $0,2$  \\
\hline
 &$D_{sJ}^*(2317)$         & 
 $D_{sJ}(2460)$             & 
 $D_{s1}(2536)$         & 
 $D_{s2}^*(2573)$           &
  $D_s^\prime$&
   $D_s^{*\prime}$ \\ 
 Decay Mode                &
       $D_S^+\pi^0$                         &
	$D_S^+\gamma,D_S^{*+}\pi^0,D^+_S\pi^+\pi^-$		        &
	  $D^*K,D\pi K$           &
	   $DK$            &
	                   &
			    \\      
\hline
 \multicolumn{7}{|c|}{Mass (MeV)}    \\
  PDG $\pm$                 &
 2317 $\pm$ 0.6                          &  
	2458.9 $\pm$ 0.9	        &
            2535.35 $\pm$   0.6       &
	   2573.5 $\pm$ 1.7           &
	                     & \\    
 FOCUS prel.  $\pm$                & 
 2323 $\pm$ 2                          & 
                        & 
            2535.1 $\pm$ 0.3                 &
            2567.3 $\pm$ 1.4    &  
                           &
	                       \\ 
			       \hline
 \multicolumn{7}{|c|}{$BR(D\pi K)/BR(D^* K)$}    \\			       
 BELLE prel.  $\pm$                & 
                    & 
                             &  
            $2.8 \pm 0.2 \pm 0.4 $\%&
               &  
                           &
	                       \\ 
			       \hline			        			        		 
 \multicolumn{7}{|c|}{Width (MeV)}    \\
 PDG $\pm$                 &
          $<$4.6                 &
	     $<$5.5              &
	    $<$2.3 \@ 90 \% cl       &
	     15 $\pm$ 5        &
	                     &
			        \\  
 FOCUS   $\pm$                &
                             & 
                              &  
               1.6 $\pm$ 1.0             &
             28 $\pm$ 5   &  
                           &
	                       \\ 				
 \hline
 \end{tabular}
\end{table}

 What is really new is a couple of results from BELLE \cite{Drutskoy05,BELLE0507030}
 presented at (northern hemisphere) 
 Summer conferences, most notably measurement of branching ratios,and observation of a
 nonresonant decay 
 of $D_{s1}^+(2536)$. BELLE have studied the decay $\bar{B^0} \rightarrow D_{sJ}(2317)^+ K^-$
 which is an interesting decays because quark content of final state is totally different
 from B meson, suggesting non-trivial decay mechanisms: W-exchange, final state interactions,
 tree diagram if the $\DsJ$ has a 4-quark structure. The new measurement improves previous
 low-statistics results. BELLE find a very large isospin breaking, namely that the rate for
 $\bar{B}^0$ decays to $D_{sJ} (2317)$ 
 is about three times larger than rate to $D_{sJ}(2460)$. BELLE also study the resonant
 structure of $D_{s1}(2536)^+ \rightarrow D^+\pi^-K^+$ decay, finding a small but non-zero
 fraction of non-resonant $D^+\pi^-K^+$ component relative to resonant  $D^{*+}K^0_s$.
 Besides, they studied the presence of an S-wave component, which may give informationon
 mixing between the two newly discovered $1^+$ states.
 I expect BELLE to report on this at this Conference.
\par
\section{ The $\DsJ$$(2632)^+$}
Following the discovery of $\DsJ$ states by BABAR and CLEO, SELEX (fixed target hadroproduction
at Fermilab with $\Sigma^-$ and $\pi^-$ beams) looked for signals in strangess-rich channels
with a charm meson, such as $D_s^+ \eta, D^0 K^+$ \cite{Evdokimov2004}. They found evidence
for a state at 2632~\MeV, with a width $\Gamma < 17$\MeV. They also found a very strong
isospin breaking, {\it i.e.,} the $D^0 K^+$
is severely depressed with respect to $D_s^+ \eta$.
\par
Given the interest of the SELEX results, quite immediately all other active charm experiments
looked for confirmation. Photoproduction experiment FOCUS looked \cite{Kutschke2632} in
$D^+K_s,D^0 K^+$, BABAR\cite{BABAR0408087} in $D_s^+\eta, D^0K^+, D^{*+}K_s$, BELLE
\cite{BELLE0507028} in $D_s^+\eta, D^0K^+$. All three experiments saw no evidence. Unless a
peculiar production mechanism related to the hyperon beam is in place here, one should
consider the SELEX evidence not confirmed. Results from hadron beam experiments (CDF/D0 at
Tevatron, and
COMPASS at CERN) would be useful to shed light and revive the case for the $D_{sJ}(2632)$.
\section{Changing Paradigma Of HQ Sectroscopy -  A Plethora of Ideas}
Needless to say, the BABAR and CLEO  discoveries spurred a plethora of theory papers. My
personal list of favourite topics sees the idea of Ref.~\cite{BARDEEN} in top position: combine HQS
and chiral invariance, form doublets by pairing $(D_s^+,D_s^{*+})$ with 
$(D^*_{s0^+}(2315),D_{s1^+}(2460))$. By applying chiral dynamics they find that the splitting
 between doublets should follow the prediction, indeed verified, 
 \beq
 \Delta M \equiv M(D^*_{s0^+}(2315))-M(D_s) ~
 M(D_{s1^+}(2460))-M(D_s^*)~ m_N/3 
 \eeq
 An interesting comment was made \cite{URALTSEV0406086} on the relative production rate of
 3/2 versus 1/2 states. 
 Sum rules predict dominance of 3/2 states (such as $D_1$ and $D_2^*$) 
 versus 1/2 states (such as broad $D_0^*$ and $D_1$). Since experimentally the opposite
 is observed, the author suggests the discrepancy be reconciled with lower mass 1/2 states,
 compatible to those found by BABAR and CLEO.\footnote{
 {\it Note added in proof -}
 Recent BELLE results \cite{Abe:2005up}   
 seem to suggest that the semileptonic decay $B\rightarrow X_c  \ell \bar \nu$
 is predominatly due to $X_c=D(L=1, j_q = 3/2)$ states. For a recent review see 
 \cite{Bigi:2005ff}.
 } 
 \par
 As for the SELEX evidence, it was noted\cite{MAIANI04} how, if the SELEX $D_{sJ}(2632)$
 state was confirmed 
 experimentally, the very strong 
 isosping breaking could be explained by a 4q  structure $[cd][ds]$.
 \par
 Reviewing the theory ideas put forward is beyond the scope of this paper,
 the interested reader can avail of several reviews, such as
 \cite{Colangelo:2004vu}. 
\section{Outlook and conclusions}
 We should be aware of the exciting era we are living, at least as far heavy quark
 spectroscopy is concerned. A lot of new results from all charm experiments active today have
 urged the need for a critical revision of the basic assumptions in the paradigma used so
 far.
 \par
 The discoveries of $\DsJ$ states by BABAR and CLEO ask for a critical revision of the HQS
 paradigma. BELLE entered the game  with confirmation of states, new decay modes, and a
 flurry of new results.
 Non-strange broad states have now been established, with FOCUS and BELLE confirming the 1998
 evidence by CLEO. The PDG average for the newly observed states sums up mass values in mild
 disagreement, more data is needed and results should come soon. 
 \par
 The intriguing evidence of
 $D_{sJ}(2632)$ by SELEX is not confirmed by FOCUS, BABAR, BELLE. We could be experiencing a
 peculiar  production mechanism connected to the strangeness-rich beam, or simply a
 statistical fluctuation. There is real opportunity for hadron beam experiments (CDF/D0 at
 Tevatron, COMPASS at CERN) to say the last word on the issue. As a lot of work is being done
 presently world-wide, we should expect a wealth of new results in plenary 
 (Mueller\cite{Muller:2005ju},
 Trabelsi\cite{Trabelsi05}, Maciel\cite{Maciel05}
 ) and parallel (Kopar\cite{Kopar05}, Poireau\cite{Poireau05}, Cumalat\cite{Cumalat05},
 Lesiak\cite{Lesiak05}) session talks. 
 \par
 Where are we going next ? Of course the list of open problems is fairly large, just to quote
 some:
 \begin{itemize}
    \item establish the non-strange broad states. In particular  all channels with neutrals
    are unobserved so far;
    \item measure the widths of $\DsJ$ states;
    \item measure the relative production of 1/2 versus 3/2 states;
    \item solve the mystery of the existence of the radial excitations.
    \item investigate the {\it Terra Incognita}: the beauty  L=1 mesons, verify the little
    data available\cite{Fabbri:1999mf}, mainly dating back to LEP, and check the validity of
    flavour symmetry, if any. 
 \end{itemize}
Most of this shopping list will be addressed by  experiments at B-factories. Hadroproduction
(CDF/D0) may contribute, as well fixed target (COMPASS at CERN). In the far future I see only
SuperBELLE as a player, after the 
cancellation of the flavour programme in the US. LHC-b seems to be ruled out by the choice of
not triggering on charm decays. PANDA \cite{BOCAHADRON} will be a major player in charmonium
spectroscopy, but seems to me problematic in taming the huge minimum bias background in the
search of charm decay verteces. As for the B spectroscopy sector, which is crucial to verify
the extent to which one can still apply flavour symmetry, it should be playground of LHC-b at
CERN. In any case, the field literally bursts with enigmas, and loads of good data are coming
in.

\begin{theacknowledgments}
  I sincerely thank the Organizers for a perfectly enjoyable meeting. I acknowledge
  enlightening discussions with  D.~Asner, T.~Barnes, I.~Bigi, G.~Boca, S.~Chung, J.~Cumalat,
  A.~Maciel,   S.~Paul, A.~Reis, K.~Seth. I thank F.de~Fazio, 
  K.Harder, U.~Karshon and T.~Matsuki for 
  comments on this paper.
\end{theacknowledgments}

%

\end{document}